\documentclass[12pt]{article}

\begin{document}


%
%

\title{Nondecoupling of Heavy Fermions and a Special Yukawa Texture
}

\author{M. N. REBELO\footnote{
On sabbatical leave from Departamento de F{\'\i}sica and Centro  de F{\'\i}sica
Te{\'o}rica de Part{\'\i}culas (CFTP),
Instituto Superior T\'{e}cnico, Av. Rovisco Pais, 1049-001
Lisboa, Portugal.  E-mail: rebelo@ist.utl.pt} \\
CERN, Department of Physics, Theory Division \\
CH-1211 Gen\`eve 23, Switzerland \\
margarida.rebelo@cern.ch}



\maketitle


\begin{abstract}
Talk based on work entitled  ``Yukawa textures, new physics 
and nondecoupling,'' done in collaboration with
G.~C.~Branco and J.~I.~Silva-Marcos,  arXiv:hep-ph/0612252, 
to appear in Phys. Rev. D.
In this work we pointed out that 
New Physics can play an important r\^ ole in rescuing some of the 
Yukawa texture zero ans\" atze which would otherwise be eliminated 
by the recent, more precise measurements of $V_{CKM}$. We have shown 
that the presence of an isosinglet vector-like quark which mixes 
with standard quarks, can render viable a particularly interesting
four texture zero Yukawa ansatz. The crucial point is the 
nondecoupling of the effects of the isosinglet quark, even for 
arbitrary large values of its mass.

\end{abstract}


\section{Introduction}
	
There have been many attempts at understanding the pattern of quark and
lepton masses as well as mixing, including the addition of family symmetries
and the introduction of special textures for the Yukawa couplings. 
The underlying idea is that such models could be derived
from grand unified models, or possibly models with extra dimensions, and
be related to special symmetries.

The recent experimental discovery
of neutrino oscillations pointing towards non-vanishing neutrino masses,
has rendered the flavour puzzle even more exciting. Nonzero neutrino masses
are evidence of physics beyond the Standard Model and have profound 
implications in the physics of the leptonic sector\cite{Mohapatra:2005wg} . 
Indeed this opens the
possibility of having leptonic CP violation both at low and high energies.

The dynamical generation of the Baryon Asymmetry of the Universe (BAU) 
is also a fundamental challenge both for particle physics and cosmology. It 
has been established that CP violation present in the Standard Model (SM) 
through the Kobayashi-Mashawa mechanism, is not sufficient to generate the 
observed BAU \cite{Gavela:1994dt}. 
As a result, the need for a viable baryogenesis provides the 
strongest argument in favour of the existence of new sources of CP violation. 
Leptogenesis \cite{Fukugita:1986hr} is  one of the most elegant 
and plausible 
scenarios for generation of BAU, specially due to the recent experimental 
evidence for a non-vanishing mass for neutrinos.

The availability of increasingly
more precise knowledge of the mixing matrices both in the quark and 
lepton sectors constitutes a great challenge to flavour models.

At present all experimental data on flavour physics and CP violation
in the quark sector are in agreement with the SM and its KM mechanism.
This is a
remarkable success, since it is achieved with a relatively small number of
parameters. Once the experimental values of $|V_{us}|$, 
$|V_{cb}|$ and $|V_{ub}|$ are
used to fix the angles $\theta_{12}$, $\theta_{23}$ and 
$\theta_{13}$ of the standard parametrization, 
one has to fit with a single parameter $\delta_{13}$,
the experimental values of $\epsilon _{K}$, $\sin ( 2\beta)$,
$\Delta M_{B_{d}}$, as well as $\Delta M_{B_{s}}$ and $\alpha$.

Recently a precise measurement of
the angle $\beta$ \cite{Aubert:2004zt} \cite{Itoh:2005ks} \cite{Abe:2005bt}
\cite{Sciolla:2005kz}
has been achieved together with the recent measurement 
of the rephasing invariant angles $\gamma$ \cite{Poluektov:2004mf}
\cite{Abe:2004mu} \cite{Abe:2004gu} \cite{Aubert:2006ia}
\cite{Poluektov:2006ia}
and $\alpha$ \cite{Aubert:2004zr},
\cite{Aubert:2005nj} \cite{Somov:2006sg} \cite{Aubert:2006af}
and the measurement of
$B_{s}^{0}$--$\bar{B}_{s}^{0}$  mixing \cite{Abazov:2006dm}
\cite{Abulencia:2006mq}.
These recent measurements  will be improved in the future. The
angle $\gamma$ plays a crucial r\^ ole in providing irrefutable 
evidence for a complex CKM matrix \cite{Botella:2005fc} 
all these  measurements place constraints 
on the size of New Physics contributions \cite{Ligeti:2004ak}
\cite{Silvestrini:2005zb} \cite{Bona:2007mn} \cite{Charles:2004jd}
\cite{Botella:2006va} \cite{Lunghi:2007ak}.

Furthermore the BaBar and Belle experiments have recently presented
evidence for $D^{0}$--$\bar{D}^{0}$  mixing \cite{Aubert:2007wf}
\cite{Staric:2007dt}
with no hint for new physics \cite{Nir:2007ac}, at least 
not yet \cite{Branco:1995us} \cite{Ball:2007yz} \cite{Golowich:2007ka}.

In spite of the
impressive results on flavour physics and CP violation in the quark sector 
there is still plenty of room for the presence of New
Physics. Furthermore it has been observed \cite{Yao:2006px} that the size of
$|V_{ub}|$ is somewhat above the range favoured by the measurement of 
$\sin ( \beta)$. Another small discrepancy is 
the fact that the central 
value of the SM prediction \cite{Misiak:2006zs} \cite{Misiak:2006ab}
for the inclusive radiative 
$\bar{B} \rightarrow  X_s \gamma$ decay, with $\bar{B} = \bar{B}^{0}$ or
$B^-$, is now more than $1 \sigma$ \cite{Haisch:2007ic}
below the experimental average \cite{Chen:2001fj}
\cite{Koppenburg:2004fz} \cite{Aubert:2006gg} \cite{Barberio:2007cr}.

In what follows we describe how the presence of an isosinglet
vector-like quark which mixes with the standard quarks can render viable
a particularly interesting four texture zero ansatz which has been 
recently ruled out by experiment \cite{Branco:2006wv}.

Isosinglet vector-like quarks play a r\^ ole very similar to righthanded
singlet neutrinos in the leptonic sector in the seesaw framework 
\cite{Minkowski:1977sc} \cite{Yanagida:1979as} \cite{glashow}
\cite{GellMann:1980vs} \cite{Mohapatra:1979ia}.
Models with vectorial isosinglet quarks together with righthanded 
singlet neutrinos
and an extra complex singlet Higgs field allow for 
a common origin for all CP violations  \cite{Branco:2003rt} 
\cite{Achiman:2004qf} \cite{Adhikary:2005fh}
including the possibility of leptogenesis.
Furthermore, these models provide a possible solution to the strong 
CP problem \cite{'t Hooft:1976up} \cite{'t Hooft:1976fv}
of the type proposed by Nelson \cite{Nelson:1983zb} \cite{Nelson:1984hg}
and Barr \cite{Barr:1984qx}
as was previously shown \cite{Bento:1991ez}.
Isosinglet quarks give rise to deviations from unitarity of the
$V_{CKM}$ matrix and Z flavour changing neutral 
currents  \cite{Branco:1986my} \cite{Nir:1990yq} \cite{Branco:1992wr}
\cite{Barenboim:1997qx} \cite{Barenboim:2001fd} \cite{Hawkins:2002qb}
producing new physics effects which may be observed in the LHC
in the near future  \cite{AguilarSaavedra:2002kr}.

\section{Four Zero Hermitian Ansatz embedded into a Model with One 
Down Vectorial Quark}

In our work we considered a specially interesting four zero Hermitian 
ansatz which had been analysed in detail in the literature 
\cite{Branco:1999nb} \cite{Fritzsch:1999ee} \cite{Roberts:2001zy}
where the  quark mass matrices $M_{u}$, $M_{d}$
are assumed to have the form:
\begin{equation}
M_{u}=\lambda _{u}\ \ K_{u}^{\dagger }\ \ \left[
\begin{array}{ccc}
0 & a_{u} & 0 \\
a_{u} & b_{u} & c_{u} \\
0 & c_{u} & 1-b_{u}
\end{array}
\right] \ K_{u}\quad ;\quad M_{d}=\lambda _{d}\ \ \left[
\begin{array}{ccc}
0 & a_{d} & 0 \\
a_{d} & b_{d} & c_{d} \\
0 & c_{d} & 1-b_{d}
\end{array}
\right]  
\label{massas}
\end{equation}
with $K_{u}=diag(e^{i\phi _{1}},1,$ $e^{i\phi _{3}})$ 
and all other parameters real. 

The Hermitian matrices given by Eq.~(\ref{massas}) give rise to leading order
to the well known texture zero relations\cite{Hall:1993ni} 
\cite{Ramond:1993kv} \cite{Barbieri:1997tu} \cite{Roberts:2001zy}:
\begin{equation}
\left| \frac{V_{ub}}{V_{cb}} \right| = \sqrt{\frac{m_u}{m_c}}
\qquad \left| \frac{V_{td}}{V_{ts}} \right| = \sqrt{\frac{m_d}{m_s}}
\qquad \left|V_{us} \right| = \left|
\sqrt{\frac{m_d}{m_s}}e^{i \phi_1} - \sqrt{\frac{m_u}{m_c}} \right| 
\label{eq4}
\end{equation}
The requirement of Hermiticity is important in order to render 
texture zero ans\" atze predictive, it was shown \cite{Branco:1988iq} 
that without this requirement several interesting zero textures 
would simply correspond to a choice of weak basis. 

It was already pointed out before  \cite{Roberts:2001zy} that the relation 
obtained for 
$\left| V_{ub}\right| / \left| V_{cb}\right|$ strongly 
disfavoured this ansatz due to the smallness of the ratio 
$\sqrt{\frac{m_u}{m_c}}$. In the meantime the experimental 
value  \cite{Yao:2006px} for
$\left| V_{ub} \right| $ went up significantly and this constraint 
became even more stringent. The determination of the ratio 
$\left| V_{td}\right| / \left| V_{ts}\right|$ has been 
significantly improved and is theoretically clean, its present
value  \cite{Yao:2006px} lies below the ratio 
$\sqrt{\frac{m_d}{m_s}}$ therefore also disfavouring this ansatz. 
Another source of difficulties lies in $\sin 2 \beta$ since ans\" atze
such like the one we are considering have no non-factorizable
phases and therefore produce too small values  \cite{Branco:2004ya} 
for the angle $\beta $. Furthermore in the framework of this ansatz the
value of $\gamma$ is very constrained and tends towards  too
large a value which may be ruled out once the experimental errors 
are reduced. 

The difficulty in accommodating $\beta $ and the ratio  
$\left| V_{td}\right| / \left| V_{ts}\right|$ could be 
avoided by assuming that there are new physics contributions to 
$B_{d}^{0}$--$\bar{B}_{d}^{0}$ and  $B_{s}^{0}$--$\bar{B}_{s}^{0}$ 
mixings. However the remaining difficulties with the extraction of 
$\left| V_{ub}\right| / \left| V_{cb}\right|$ and 
$\gamma $ would not be solved 
since they are unaffected by the presence of new physics in the
mixing.

In order to overcome these problems we embedded the above ansatz
into an extension of the SM with one additional $Q=-1/3$ isosinglet 
vector-like quark.
Therefore we analysed the larger ansatz obtained with the same choice for 
$M_{u}$ and $M_d$ placed in the new $ 4 \times 4 $ down quark
matrix  ${\cal M}_d$ in the following way:
\begin{equation}
{\cal M}_d = \left(\begin{array}{ccccc}
  &  &   & | & 0  \\
  & M_d &  & | & 0  \\
 &  &  & | & 0  \\
- & - & - &  | & - \\
  & M_D & & | & H  \\  
\end{array}\right)
\label{pra}
\end{equation}
With only one extra vectorial quark   $M_D$ is
a $1 \times 3$ matrix and $H$ a single entry. Since  the mass terms 
$M_D$, $H$ are $SU(2) \times U(1)$ invariant  they can be much 
larger than the electroweak scale. We may assume that there is a family 
symmetry which leads to the four texture zero ansatz, in the $3 \times 3$
quark mass matrices involving standard quarks which is softly 
broken by the terms $M_D$ and  $H$.
The matrix ${\cal M}_d$ is diagonalized by the usual 
bi-unitary transformation:
\begin{equation}
U^{\dagger}_{L}{\cal M}_d U_{R} =  \left(\begin{array}{cc}
{\overline m} & 0 \\
0 & {\overline M} 
\end{array}\right)
\label{tttt}
\end{equation}
where ${\overline m} =$ diag $(m_d, m_s, m_b )$ and  
${\overline M} $ is the heavy quark mass. One can write $U_L$ in block form,
\begin{equation}
U_L  =  \left(\begin{array}{cc}
 K  & R \\
 S & T 
\end{array}\right)
\label{krst}
\end{equation}
where $K$ is the usual $ 3 \times 3$  $V_{CKM}$ matrix. $U_L$ is the matrix 
that diagonalizes  ${\cal M}_d {\cal M}^{\dagger}_d$, and the 
following relations can be readily derived \cite{Bento:1991ez}
in the limit $ M_D, H > > {\cal O} (M_d) $
\begin{eqnarray}
\overline{M} ^2 \simeq (M_D {M_D}^\dagger + H^2)  \equiv M^2 \\
\overline{m} ^2 \simeq K^{\dagger} m_{eff}  m^{\dagger}_{eff} K
\label{mem}
\end{eqnarray}
with 
\begin{equation}
 m_{eff}  m^{\dagger}_{eff} \simeq M_d {M_d}^\dagger -
\frac{(M_d {M_D}^\dagger M_D \  {M_d}^\dagger)} {M^2}
\label{eee}
\end{equation}
Note that $K$ is the mixing matrix connecting standard quarks
and has small deviations from unitarity given by
$K^{\dagger} K = 1 - S^{\dagger} S$, with:
\begin{equation}
 S \simeq - \frac{M_D M^\dagger_d K}{M^2} \left( 1 + 
\frac{\overline{m}^2} {M^2} \right)
\label{sss}
\end{equation}
It is the fact that second term in Eq.~(\ref{eee}) may be of a 
magnitude similar to that of  $M_d {M_d}^\dagger$ that makes 
it possible to rescue the four texture ansatz discussed before.

The present experimental data can be well reproduced 
 \cite{Branco:1999tw}, \cite{Roberts:2001zy}
from the following Froggatt-Nielsen pattern \cite{Froggatt:1978nt}
for  $m_{eff}$:
\begin{equation}
m_{eff} \sim m_b 
\left(\begin{array}{ccc}
  0 & \overline{\varepsilon}^3  &   \overline{\varepsilon}^4 \\
   \overline{\varepsilon}^3  &  \overline{\varepsilon}^2 &
   \overline{\varepsilon}^2    \\
   \overline{\varepsilon}^4 &  \overline{\varepsilon}^2 & 1   \\
\end{array}\right),
\label{exp}
\end{equation}
with $ \overline{\varepsilon} \simeq 0.2$ together with a similar
pattern for the up sector. However due to the different hierarchies 
of the quark mass matrices the $V_{CKM}$ matrix is specially 
sensitive to the down sector and we can still reproduce the
present experimental data with the form chosen for $M_u$ in 
Eq.~(\ref{massas}).
   
The specific patterns in terms of  $ \overline{\varepsilon}$ 
chosen for  ${\cal M}_d$ in order to render viable this Yukawa texture
are discussed in our original work \cite{Branco:2006wv} where a
numerical example can also be found.

We would like to stress the important fact that a sizeable effect in
$m_{eff}  m^{\dagger}_{eff}$ can be obtained even in the limit of
an extremely heavy vectorial quark. For  $M_D M^{\dagger}_D$ and $H^2$  
of the same order of magnitude Eqs.~(\ref{eee}) and (\ref{sss}) show that
the desired effect can be obtained in $m_{eff}  m^{\dagger}_{eff}$
and at the same time, deviations from unitarity of the $3 \times 3$ 
$V_{CKM}$ matrix are suppressed due to the smallness of the entries in $S$.

\section*{Acknowledgments}

The author thanks the organizers of the CTP 2007 Symposiun 
on Sypersymmetry at LHC for the warm welcome at the 
Centre for Theoretical Physics at The British University in Egypt
and the stimulating scientifical environment provided. 
This work was partially supported by Funda\c c\~ ao para a 
Ci\^ encia e a  Tecnologia (FCT, Portugal) through the projects
POCTI/FNU/44409/2002, PDCT/FP/63914/2005, PDCT/FP/63912/2005 and
CFTP-FCT UNIT 777 wich are partially funded through POCTI 
(FEDER).


\end{document}